%% file: main.tex
\documentclass[11pt,twocolumn,a4]{article}
\usepackage[top=2.1cm, bottom=2.1cm, left=1.5cm, right=1.5cm]{geometry}
\usepackage{pgf}
\usepackage{pgfpages}
\usepackage{fancyhdr}
\usepackage{amssymb}
\usepackage{amsmath}
\usepackage{textcomp} 
\usepackage{xspace} 
\usepackage{pifont} 
\usepackage{stmaryrd} 
\usepackage{mdwlist} 

\pgfpagesdeclarelayout{boxed}
{
  \edef\pgfpageoptionborder{0pt}
}
{
  \pgfpagesphysicalpageoptions
  {%
    logical pages=1,%
  }
  \pgfpageslogicalpageoptions{1}
  {
    border code=\pgfsetlinewidth{0.4pt}\pgfstroke,%
    border shrink=\pgfpageoptionborder,%
    resized width=.935\pgfphysicalwidth,%
    resized height=.935\pgfphysicalheight,%
    center=\pgfpoint{.5\pgfphysicalwidth}{.5\pgfphysicalheight}%
  }%
}

\pgfpagesuselayout{boxed}

\begin{document}
\title{From abstract modelling to remote cyber-physical integration/interoperability testing}

\author{%
Maria Spichkova, %
Heinrich Schmidt %
and
Ian Peake 
\\ 
Computer Science and IT, RMIT University, Melbourne, AUSTRALIA
\\ 
Email: \{Maria.Spichkova,Heinz.Schmidt,Ian.Peake\}@rmit.edu.au
}

\lhead{}
\chead{}
\rhead{\small{Improving Systems and Software Engineering Conference incorporating}\\
\small{SEPG$^{SM}$ Asia-Pacific Conference 2013}\\
\small{Melbourne, September, 2013}%
}

\lfoot{}
\cfoot{}
\rfoot{\thepage}

\date{}

\maketitle

\begin{abstract} 
An appropriate system model gives developers a better overview, and the ability to fix more inconsistencies more effectively and earlier in system development, reducing overall effort and cost. 
However, modelling assumes abstraction of several aspects of the system and its environment, and this abstraction should be not overlooked, but properly taken into account during later development phases.  
This is especially especially important for the cases of remote integration, testing/verification, and manufacturing of cyber-physical systems. 
For this reason we introduce a development methodology 
for cyber-physical systems (CPS) with a focus on the abstraction levels of the system representation, 
based on  the idea of refinement-based development of complex, interactive systems. 
\end{abstract}

\newcommand{\Prop}{\mathbb{PROP}}
\newcommand{\PropL}[1]{\mathbb{LPROP}^{#1}}
\newcommand{\PropA}[1]{\mathbb{ABSTR}^{#1}}
\newcommand{\PropAK}[1]{\mathbb{ABSTR_{KNOW}}^{#1}}
\newcommand{\EnvAsm}[1]{\mathbb{ENV_{ASM}}^{#1}}

\input{intro}

\input{related} 
\input{abstraction}

\input{aspects} 
\input{testing}
\input{scenarios}

\section{Conclusion}
In this paper we have discussed our vision of the modelling levels for cyber-physical systems, 
how this idea correlates with the ideas of remote cyber-physical integration/interoperability testing in a virtual environment, to minimise the overall production effort,
which system's aspects should be treated more carefully and more particularly, and
which scenarios can be used to develop a practical  methodology for the cases of cyber-physical system manufacturing that is distributed over different locations. 

\bibliographystyle{IEEEtran}

\end{document}

%% file: intro.tex
\section{Introduction}

Generally in engineering and production, the key question is how to reach engineering goals in a way that minimises the overall effort. 
A large part of the answer to this question in system engineering is in choice of an appropriate design process as well as appropriate abstraction levels during each design stage.
In state-of-the-art industrial development, quality assurance is performed by extensive testing of generated code and of the real system which is physically present for testing. 
However, testing can only demonstrate the absence of errors for exemplary test cases, 
not the correctness of the system with respect to its requirements. 

In contrast to testing,
{\em verification} delivers a correctness proof for critical properties,
but requires significant effort,
especially if we refer to verification of the code,
which is typically more complex than
verification on the model level,
which deals with abstraction from the real world,
and we need only demonstrate correctness of the system model and not of the system at the level of code
(or even in principle at the physical level).
In some cases, even after verifying certain properties on the modelling level,
inconsistencies can still remain in the real system, as elegantly stated by Donald E. Knuth's famous saying: 
``Beware of bugs in the above code -- I have only proved it correct, not tried it.'' 
Thus,  neither  verification on the modelling level nor testing alone should be trusted in isolation of each other. 
To minimise overall effort while ensuring required system properties we need to combine these techniques. 

Modelling a system assumes abstraction of several aspects of the system and its environment,
which implies that no model can fully represent and substitute for a real system,
however, in many cases we do not need the whole representation of a system, but only the representation of some its parts relevant to a concrete purpose. 
Thus, an appropriate model is always developed by taking into account the question of what the model should serve for:
\begin{itemize}
\item 
On which aspects should we focus to cover all properties which are important with respect to the current purpose of the model?
\item
From which aspects should we abstract to obtain better readability and scalability of the model?
\item
Which aspects can be represented as parameters to obtain a model which is easy to reuse and to extend?
\end{itemize}
From this perspective,
a further interesting question is whether remote cyber-physical integration/interoperability testing,
as the next phase in the development process,
can or should have any influence on the elaboration of an abstract model and its analysis on the logical level.
(We assume hereafter discrete representations of time and sensor data.)
Can we take advantage of abstraction of information about the further development process?
Indeed, can we even defer the decision of whether some integration and testing steps will be done remotely for the future?
Otherwise, which modelling aspect should be taken into account already on the logical level if the development process includes remote integration and testing?

This question is especially complicated, if we focus on cyber-physical system production that is distributed over different locations. 
The real development process never fits to the waterfall development model;
single components and connections between them are optimised and corrected in the development loop a number of times after the system's testing/verification, 
until the deployed system fulfils all the requirements.
In the case of system manufacturing at different places,
transport from component development locations to integration and deployment locations can significantly increase the whole development costs.
If system's components are bulky or heavy, this may also delay optimisation and correction.
Moreover, in traditional engineering processes,
frequently duplication of specialised components and equipment as well as duplication of corresponding skilled staff in multiple locations is needed.
One partial solution to this problem is a virtual interoperability testing approach~\cite{vitelab},
where the development process is extended by an additional level of abstraction --- the level of remote virtual system representation.
Cyber-physical components are remotely embodied to each other and integrated with the aid of a virtual interoperability test lab (VITELab),
consisting of sensors, actuators, networks and simulation.
The aim of a VITELab approach is to reduce costs associated with interoperability testing at integration and commissioning phases in software-intensive automation application. 
In this paper we present the VITELab development model as whole, starting from an abstract logical view on a system.

%% file: related.tex
\section{Related Work}

One of the noted approaches on system modelling and simulation is  Modelica~(cf.\ \cite{modelica2008}, \cite{Fritzson2004Modelica}, \cite{modelica2013vehicle})  which covers 
modelling, simulation (also in the sense of distributed real-time simulation) and verification of discrete control components.
Modelica is mostly object-oriented
and its latest extensions allow also modelling of system requirements~\cite{modelica2013req} as well as 
simulation of technical and physical systems~\cite{Fritzson2011Modelica}.
Modelling theories for distributed hybrid system as SHIFT~\cite{Deshpande97shift:a} and R-Charon~\cite{KratzSPL06} guarantee complete simulation and compilation of the models,
but do not support verification or analysis of the system on the modelling level.
The same limitation applies to UPPAAL~\cite{DBLP:conf/sfm/BehrmannDL04} and PHAVer~\cite{Beek06syntaxand}, which provide the simulation, but a limited verification with restricted dynamics and only for small fragments.

In traditional development of embedded systems (cf., e.g., \cite{ES_Berger}), 
the system is usually separated into software and hardware parts as soon as possible, at an early stage of the development process,
however, in an abstract level of modelling it may be easier in many cases to ignore the difference in the nature of components. 
\cite{Sapienza2690} and \cite{Spichkova_Campetelli2012} independently suggested to use a platform-independent design in the early stages.
The approach presented in \cite{Sapienza2690} introduces the idea of pushing
hardware- and software-dependent design to a stage that is as late as possible, 
however, the question of the current practical and fundamental limitations of logical modelling in comparison to cyber-physical testing, 
is still not completely answered, especially for the case of digital representability of cyber-physical systems (CPS). 

In comparison to \cite{Sapienza2690}, 
the focus of the first author's previous work~\cite{Spichkova_Campetelli2012}  
was on reutilisation of the existing methodology for the development of software systems for application within the cyber-physical
domain to benefit from the advantages these techniques have shown.
More precisely, the work focused on reusing the generalisation of two methodologies, both elaborated according to the results of three case studies motivated and supported by DENSO Corporation and Robert Bosch GmbH. 
However the question, how deep can we go on the modelling of cyber-physical systems on the logical level is still open in both approaches.
From our point of view the answer strongly depends on the concrete domain of system application. 

The idea of early analysis of critical system faults has the aim to identify faults which mutate the safety critical behaviour of the system,
and to identify test scenarios which can expose such faults from the model on the logical level,
i.e. by generation of tests (both for real system and its model) from formal specifications or from the CASE tool model (cf., e.g., \cite{Daggupta2012tests,Broy2005testing,pretschner200510}). 
It has certain limitations due the abstract nature of the formal model serving as a base for the test generation as well as 
an underlying assumption of existence of a precise formal model of the system being developed, however, even taking into account these limitations and assumptions, 
it allowed automatisation of test case design and makes the whole design process more stringent. 
Test case generation can be considered the heart of testing, and model-based testing enables higher test coverage on
the logical level by generating test cases, 
and the focus of our work is to determine how the gap between the abstraction levels can influence test coverage of a real system
and how to extend the model-based testing by alerts on the system's aspects which are (possibly) uncovered by the generated tests.

Another important aspect of cyber-physical systems is that they are software-intensive.%
For general approaches on software-intensive systems we refer to \cite{SISengineering}. 

Challenges for cyber-physical systems such as security, error protection and timing analysis as well as verification and system optimisation were discussed in~\cite{CPSchallenges}
and~\cite{Cheng_synthesis}. 
There are many approaches on mechatronic/cyber-physical systems, however, most of them omit abstract logical level of the system representation and lose the advantages of the abstract representation. 
For instance, the work presented in~\cite{Vogel-Heuser_IECON} defines an extensive support to the components communication and time requirements, while the model discussed in~\cite{IEEE_INDIN_2011} proposes a complete model of the processes with communication. 
Nevertheless, representing a real, both in a sense of physical representability and real size, system by a single abstraction level  could be a disadvantage in the project of a cyber-physical system, where experts of different domains should be able to cooperate and work to different views and abstraction levels of the system.

%% file: abstraction.tex
\section{Levels Of Abstraction}

Any requirements specification of a system can be seen as an abstraction of the system (cf. also \cite{spichkova2008refinement,ArchReqDecRef}): 
collecting a list of system requirements we get an abstract system description that is in most cases a black-box view, or, more precisely, 
it must be a black-box view in an ideal case, but in industrial requirement specification we get in many cases a mix of black- and glass-box view,
i.e. a mix of requirement and architecture specification that must be separated to get a clear and readable system specification. 
In some cases (using an existing requirements, i.e. black-box, specification) we also need to represent some properties in more abstract way for better supporting of maintainability and modularity aspects. 

For many systems of interest in industrial automation uncertainty plays a vital role, for example arrival time of messages or probability distribution for failures or usage etc. In cyber-physical systems many exogenous events that control or trigger behaviour are stochastic. 
For quantitative properties like worst case execution time,  latency, reliability etc the statistical nature of some of the interface to the environment will have to be pushed to subsystem interfaces between systems that would otherwise be considered 'internal'. 
For all these reasons, for cyber-physical systems, we need a more flexible way of connecting ports, groups, sheaths and algebras of gates stochastically to other gates both spatially and temporally.

Speaking about abstraction, i.e. about reduction of complexity by reducing details, we should take into account 
whether we can mark some system properties or parts  as too concrete for the current specification layer and 
omit them for keeping the model more readable and abstract, i.e., to omit  some system's properties and  components, we have to cheek that whether we loose any important information about the system, on this level of abstraction or in general. 
Especially important is to analyse (in general as well as for a concrete case), which kind of system's aspects and assumptions on the logical level impact most on fidelity of the model with respect to the real system.\\
~\\

\begin{figure}[ht]
\begin{center}
\includegraphics[scale=0.5]{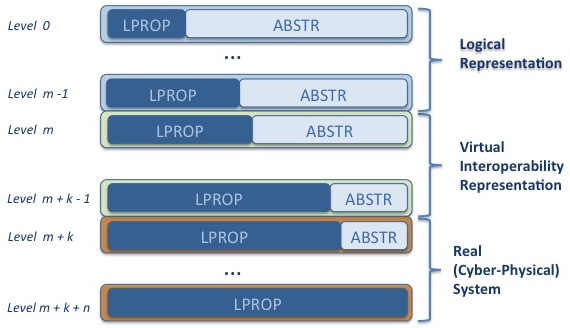}
\end{center}
\caption{Abstraction Levels}
\label{fig:sbstrLevels}
\end{figure}

\newpage
If the information is not important on the current level, it could influence on the overall modelling result after some refinement steps, i.e. at more concrete levels that are more near to the real system in the physical world. Therefore, specifying system we should make all the decisions on abstraction in the model transparent and track them explicitly -- in the case of contradiction between the model and the real system this allows to find the problem easier and faster. 
In general, we can say that any system $S$ can be completely described by the corresponding  set $\Prop(S)$ of its (cyber-physical) properties. On each level $l$ of abstraction we can split it into two subsets: set  $\PropL{l}(S)$ of the properties reflected at this level of abstraction, and set $\PropA{l}(S)$ of properties from which we abstract at this level, knowingly or unknowingly. 

We denote by $\PropAK{l}(S)$ the properties of the system from which we abstract intentionally and which we aim to track during system development, 
$\PropAK{l}(S) \subseteq \PropA{l}(S)$. 
For any abstraction level  $l$ the following holds:\\
 $\PropL{l}(S) \cup \PropA{l}(S) = \Prop(S)$ and\\
  $\PropL{l}(S) \cap \PropA{l}(S) = \emptyset$. 
  
Thus, with each refinement step we move some part of system's properties from the set $\mathbb{ABSTR}$ to the set $\mathbb{LPROP}$, 
and in some sense the set $\mathbb{ABSTR}$  represent the termination function for the modelling process: 
in the case $l$ corresponds to the real representation of the system, we get $\PropL{l}(S) = \Prop(S)$ and $\PropA{l}(S) = \emptyset$ (cf. also Figure~\ref{fig:sbstrLevels}). 

\begin{figure}[ht]
\begin{center}
\includegraphics[scale=0.49]{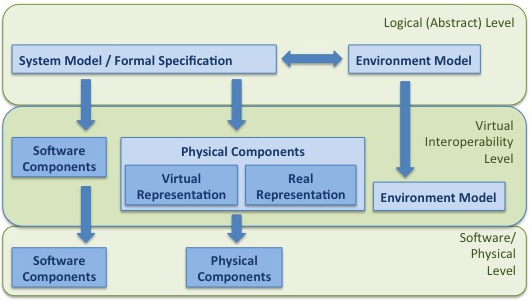}
\end{center}
\caption{Cyber-Physical Systems: Generalised Development Methodology}
\label{fig:process}
\end{figure}

On some level $m$ we need to switch from the pure abstract (logical) representation of the system to a cyber-physical one, but during a number of refinement steps we 
test (and refine) the system or component using a virtual environment, and then continue with testing in a real environment (cf. Figure~\ref{fig:process}). 

Moreover, on the logical level $l$ we use a number of assumptions on environment of a system $S$ and similar of each single logical component. 
We denote this set of assumptions by $\EnvAsm{l}$. 
 
A basic assumption is that one does not have perfect knowledge of requirements:
testing is not only to reveal/exclude bugs which arise in (non-automated) refinements (as in verification)
but also to evaluate prototype (un)suitability which may arise from misunderstood requirements (as in validation).
Moreover in practice we view the abstraction levels as corresponding to stages in an imperfect process rather than views which are kept complementary and consistent.
Thus, in comparison to the sets $\PropAK{l}$, it is unrealistic to expect monotonicity between the number $l$ and the cardinality of the set $\EnvAsm{l}$: 
some assumptions on the environment could become weaker or unnecessary with the next refinement step,
but for some assumptions stronger versions may be needed or the system can require some new assumptions in order to fulfil all its properties.
However, it is important to trace the changes of $\EnvAsm{}$ on each level of modelling to find out which properties of the model on which levels should be re-verified,
if on some refinement step $l+1$ a contradiction between the $\EnvAsm{l}$ and the real targeting environment will be found.

Thus, the collected assumption should be checked during the testing phase, and if something is missed or incorrect, the model should be changed accordingly to the results of the testing. In some cases, this validation could be done remotely to save testing and integration costs.

%% file: aspects.tex
\section{Special Aspects Of CPS}

Before switching from the logical level to the level of the virtual interoperability representation and testing (or \emph{VITE level} for short), where everything which built an ``environment'' to our system is represented virtually, we should turn from the abstract design to the hardware- and software-dependent design first, because (a part of) the system components to test on the VITE level should be real on this stage.

Logic itself assumes a certain level of abstraction, 
and on the level of the abstract (logical) representation we have, both for cyber and for physical components of the system, discrete representation of 
time, sensor data, all the signal and data flows, etc., 
where for the real cyber-physical system we have analog representation of time, but representation of signals and sensor data might be analog or discrete depending on their real nature. However, verification by model checking or theorem proving requires discretisation of many aspects, especially of time. 
On the other hand, on certain level of abstraction,  it is possible to switch from 
the  continuous time representation to the digital one without loosing the essential properties of the represented system~\cite{DigitalClocks}.  
For this reason we have introduced in out previous work~\cite{spichkova2012time} an approach on abstract modelling of timing aspects for the system representation on logical level.

Speaking about the term ``physics" in the case of cyber-physical systems, first of all a suitable notion of space and time is needed, including some freedom for modellers to choose a space-time coordinate system for information processing.  
Furthermore, it is importantly to examine the  information flow not only in time but also in space, because the space aspect may influence in this case not only on the delays (and be modelled solely by timing aspects) but also additional information about the environment and the interaction between system and the environment as well as between subcomponents of the system. This point is crucial if we want to reduce costs of interoperability testing at integration and commissioning phases in software-intensive automation application by introducing the remote cyber-physical integration/interoperability testing within virtual environment.

Thus, we can define cyber-physical systems as  the union of subsystems of two kinds, cyber-logical and cyber-physical. 
In such a way we can separate from each other the following three aspects:
\begin{itemize}
\item 
intrinsic/endogenous (embedded/reactive or even adaptive/learned) behaviour/control,
\item
extrinsic/exogenous (remote/forced) control,
\item
behaviour/control that emerges from the coordination or constraints that are the interplay of endogenous and exogenous constraints 
imposed on the degrees of freedom of the ``uncontrolled' subsystem when many subsystems are coordinated; 
typically this is a limit process with $n\rightarrow \infty$ where n is the number of systems of the same type or supertype.
\end{itemize}
In that way we can obtain clarity what it means to simulate a purely cyber-logical (sub)system entirely on supercomputers while another (sub)system is entirely cyber-physical and interacting with the the former.
 

%% file: testing.tex
\section{Remote Interoperability Testing (VITElab)}

The situation becomes even more complicated if we take into account the problem of {\em remote cyber-physical integration/interoperability testing in a virtual environment}
(or {\em virtual interoperability testing} for short). 
In the general case this involves the co-ordinated testing of several cyber-physical systems which are intended to be transported and integrated in a single physical location as components of a single cyber-physical system.
Ideally such testing would require infrastructure providing:
(i) the co-embodiment of each component to all the others by virtue of specialised sensors, actuators and a network connection;
(ii) the cyber-simulation of the full system in a realistic simulated environment.
We refer to the corresponding test environment as a virtual interoperability test lab ({\em VITELab}).
In the traditional development process, interoperability testing of a system as whole is deferred until system is integrated at the dedicated site.
In a VITElab project~\cite{vitelab}, interoperability testing is performed early and remotely,
for example while cyber-physical components are in the prototyping stage i.e. on the workbench: 
individual components (e.g., robots, manufacturing cells),
are connected in a suitable virtual environment, without being deployed at the same place physically.
Successful testing could significantly reduce the well-documented costs arising from discovery of design faults after implementation.

Remote integration and testing mean that even an integration and testing phase of a real system assumes a certain level of abstraction: 
the network, the virtual environment and the remote embodiments may be abstractions themselves, however, this level of abstraction includes real physical components of the system (in the case of the VITElab project, e.g. real robots, plant, etc.) and much more information about the network, environment and embodiments. 
This implies that the results of system analysis on this level are much more precise and less abstract in comparison to the logical level,
but the analysis itself is more expensive than an analysis on the logical level, especially if we would like not only to test the system but also verify a number of its properties. 
Nevertheless, an abstract model can give us the possibility to identify
\begin{itemize}
\item 
a number of problems and inconsistencies on the early stage of system development and verify especially important system's properties before the real system is build and integrated;
\item
possible weak points in the system (such as some timing properties, feature interactions, component dependancies, etc) which we should focus on, during the testing phase. 
\end{itemize}

%% file: scenarios.tex
\section{Future Work} 
To develop a practical  methodology for the cases of cyber-physical system manufacturing that is distributed over different locations, 
we suggest to analyse two possible scenarios of virtual integration. 
General assumptions for both scenarios are as follows:  
\begin{itemize}
\item 
 the system to develop should consist of  $n$ identical robots (in the simplest case $n = 2$) which can interact with each other to solve the problem;
\item
interaction between robots is a subject to certain assumptions/limitations, e.g. we exclude the situation where two robots should carry together some load too heavy for a single robot. 
\end{itemize}

\noindent
\textbf{Scenario 1. Space-division multiplexing of robots during simulation:}
Robots are located in different places during the phase of  virtual interoperability testing. 

\noindent 
\textbf{Scenario 2.  Time-division multiplexing of robots during simulation:}
Only one robot is available on the test phase where we need to simulate an interaction between a number of robots.  

\noindent
The crucial question in both scenarios is what is the best way to simulate the system. 
Especially interesting point is here interaction between robots, which can have physical nature, for example, 
\begin{itemize}
\item 
one robot picks up some load from another robot;
\item 
two robots operate in one room, and collision should be excluded in the correct behaviour of the robots, in particular, collision with people, if they can enter this room (hazard situation).
\end{itemize}
Thus, in order to develop a practical methodology,  we also have to answer the following questions:
\begin{itemize}
\item 
Which exactly assumptions or limitations on the interaction do we have for each particular scenario? 
Can we derive them from the  
logical model(s) where assumptions are made explicit?
\item
Do we need on logical level different models for each of the scenarios or we can proceed with a single model? 
Should these models (this model) be different from the case where all the robots are physically present on the same place during the simulation 
like in traditional testing?
\item
Can we determine on the logical level which system information can be capsuled in parameters?
\end{itemize}
These challenging questions gives a direction to our future work on the abstract modelling within virtual interoperability testing approach. 
Another interesting direction is analysis of the tracing for assumptions on the system's environment (chains of the sets $\EnvAsm{}$) as well as tracing for the abstraction from system's properties (chains of the sets $\mathbb{ABSTR_{KNOW}}$).

%% file: main.bbl
\begin{thebibliography}{10}
\bibitem{vitelab}
{Virtual Interoperation Testing Laboratory (VITElab)}, RMIT University,
  Melbourne, Australia.

\bibitem{modelica2008}
U. Donath, J. Haufe, T. Blochwitz, and T. Neidhold, ``{A new Approach for
  Modeling and Verification of Discrete Control Components within a Modelica
  Environment},'' in International Modelica Conference, 2008.

\bibitem{Fritzson2004Modelica}
P. Fritzson, \emph{Principles of Object-Oriented Modeling and Simulation with
  Modelica 2.1}.
   Wiley-IEEE Computer Society Press, 2004.

\bibitem{modelica2013vehicle}
A. Anderson and P. Fritzson, ``{Models for Distributed Real-Time Simulation in
  a Vehicle Co-Simulator Setup},'' in \emph{{Proceedings of the 5th
  International Workshop on Equation-Based Object-Oriented Modeling Languages
  and Tools}}, H. Nilsson, Ed.  Linkoping
  University Electronic Press, 2013.

\bibitem{modelica2013req}
A. Tundis, L. Rogovchenko-Buffoni, P. Fritzson, and A. Garro, ``{Modeling
  System Requirements in Modelica: Definition and Comparison of Candidate
  Approaches},'' in \emph{{Proceedings of the 5th International Workshop on
  Equation-Based Object-Oriented Modeling Languages and Tools}}, H. Nilsson,
  Ed.  Linkoping University Electronic
  Press, 2013.

\bibitem{Fritzson2011Modelica}
P. Fritzson, \emph{Introduction to Modeling and Simulation of Technical and
  Physical Systems with Modelica}. 
  Wiley-IEEE Computer Society Press, 2011.

\bibitem{Deshpande97shift:a}
A. Deshpande, A. G\"oll\"u, and P. Varaiya, ``{Shift: A Formalism and a
  Programming Language for Dynamic Networks of Hybrid Automata},'' 1997.

\bibitem{KratzSPL06}
F. Kratz, O. Sokolsky, G. J. Pappas, and I. Lee, ``{R-Charon, a Modeling
  Language for Reconfigurable Hybrid Systems},'' 
  in \emph{Hybrid Systems: Computation and Control}, LNCS, vol. 3927, 2006, pp.
  392--406.

\bibitem{DBLP:conf/sfm/BehrmannDL04}
G. Behrmann, A. David, and K. G. Larsen, ``{A Tutorial on Uppaal},'' in
  \emph{SFM}, 2004, pp. 200--236.

\bibitem{Beek06syntaxand}
D. A. V. Beek, K. L. Man, M. A. Reniers, J. E. Rooda, and R. R. H. Schiffelers,
  ``{Syntax and consistent equation semantics of hybrid Chi},'' in
  \emph{Journal of Logic and Algebraic Programming}, 2006, pp. 129--210.

\bibitem{ES_Berger}
A. Berger, \emph{Embedded Systems Design: An Introduction to Processes, Tools,
  and Techniques}.  CMP Books, 2002.

\bibitem{Sapienza2690}
G. Sapienza, I. Crnkovic, and T. Seceleanu, ``Towards a methodology for
  hardware and software design separation in embedded systems,'' in \emph{In
  Proc. of the Seventh International Conference on Software Engineering
  Advances}.  2012,
  pp. 557--562.

\bibitem{Spichkova_Campetelli2012}
M. Spichkova and A. Campetelli, ``Towards system development methodologies:
  From software to cyber-physical domain,'' in \emph{International
  Workshop on Formal Techniques for Safety-Critical Systems}, 2012.

\bibitem{Daggupta2012tests}
A. Hazra, P. Ghosh, S.G. Vadlamudi, P.P. Chakrabarti, and P. Dasgupta,
  ``Formal methods for early analysis of functional reliability in
  component-based embedded applications,'' \emph{Embedded Systems Letters},
  vol. 5, no. 1, pp. 8--11, 2013.

\bibitem{Broy2005testing}
M. Broy, B. Jonsson, J.-P. Katoen, M. Leucker, and A. Pretschner,
  \emph{Model-Based Testing of Reactive Systems: Advanced Lectures}.  Springer, 2005.

\bibitem{pretschner200510}
A. Pretschner and J. Philipps, ``{Methodological Issues in Model-Based
  Testing},'' \emph{Model-Based Testing of Reactive Systems}, pp. 181--291,
  2005.

\bibitem{SISengineering}
M. H\"{o}lzl, A. Rauschmayer, and M. Wirsing, ``Software-intensive systems and
  new computing paradigms,'' M. Wirsing, J.-P. Ban\^{a}tre, M. H\"{o}lzl, and
  A. Rauschmayer, Eds.  Springer, 2008,
  ch. Engineering of Software-Intensive Systems: State of the Art and Research
  Challenges, pp. 1--44.

\bibitem{CPSchallenges}
F. Mueller, ``{Challenges for Cyber-Physical Systems: Security, Timing Analysis
  and Soft Error Protection},'' in \emph{{National Workshop on High Confidence
  Software Platforms for Cyber-Physical Systems: Research Needs and Roadmap}}, 2006.

\bibitem{Cheng_synthesis}
A. M. K. Cheng, ``Synthesis, verification, and optimization of cyber-physical
  systems.''

\bibitem{Vogel-Heuser_IECON}
B. Vogel-Heuser, F. S., T. Werner, and C. Diedrich, ``Modeling network
  architecture and time behavior of distributed control systems in industrial
  plant,'' in \emph{37th Annual Conference of the IEEE Industrial Electronics
  Society}, ser. IECON, 2011.

\bibitem{IEEE_INDIN_2011}
T. Hadlich, C. Diedrich, K. Eckert, T. Frank, A. Fay, and B. Vogel-Heuser,
  ``Common communication model for distributed automation systems,'' in
  \emph{9th IEEE International Conference on Industrial Informatics}, ser. IEEE
  INDIN, 2011.


\bibitem{spichkova2008refinement}
M. Spichkova, ``Refinement-based verification of interactive real-time
  systems,'' \emph{ENTCS}, vol. 214,
  pp. 131--157, 2008.
  
\bibitem{ArchReqDecRef}
M. Spichkova.
\newblock {\em {Architecture: Requirements+ Decomposition+ Refinement}}.
  \newblock In {\em{Softwaretechnik-Trends 31 (4)}}, 2011.
 

\bibitem{DigitalClocks}
T. Henzinger, Z. Manna, and A. Pnueli, ``What good are digital clocks?'' in
  \emph{Proc. of the 19th International Colloquium on Automata, Languages
  and Programming}. Springer, 1992, pp.
  545--558.

\bibitem{spichkova2012time}
Spichkova, ``{Towards Focus on Time},'' in \emph{{12th International Workshop
  on Automated Verification of Critical Systems (AVoCS'12)}}, 2012.

\end{thebibliography}
